\documentstyle[prl,aps,preprint]{revtex}

\tightenlines

\begin{document}
\draft
\title{Entropy using Path Integrals for Quantum Black Hole Models}
\author{{\small O.\ Obreg\'{o}n \thanks{%
E-mail: octavio@ifug3.ugto.mx}, M. Sabido\thanks{%
E-mail: msabido@ifug3.ugto.mx}, and V. I. Tkach\thanks{%
E-mail: vladimir@ifug1.ugto.mx. }}}
\address{{\ Instituto de F\'\i sica de la Universidad de Guanajuato, }\\
Apartado\\
Postal E-143 Le\'{o}n, Gto., M\'{e}xico}
\maketitle

\begin{abstract}
Several eigenvalue equations that could describe quantum black holes have
been proposed in the canonical quantum gravity approach. In this paper, we
choose one of the simplest of these quantum equations to show how the usual
Feynman's path integral method can be applied to obtain the corresponding
statistical properties. We get a logarithmic correction to the
Bekenstein-Hawking entropy as already obtained by other authors by other
means.
\end{abstract}



\vskip -.5truecm

\newpage

In the early seventies, in his insightful work Bekenstein \cite{ref1}
proposed the quantization of a black hole. He suggested that its surface
gravity is proportional to its temperature and that the area of its event
horizon is proportional to its entropy. In his remarkable work he
conjectured that the horizon area of non-extremal black holes plays the role
of a classical adiabatic invariant. He concluded that the horizon area
should have a discrete spectrum with uniformly spaced eigenvalues. Another
result at that same time, was the discovery of the mechanical laws in the
framework of general relativity, that govern non-extremal black holes by 
\cite{ref2}. These laws have a striking analogy with those of
thermodynamics. This similarity was not well understood until Hawking, in
his seminal work \cite{ref3} a year later, discovered that black holes
evaporate, they radiate as black bodies with a temperature proportional to
the surface gravity. One can then argue that the laws of black hole
mechanics have a real thermodynamical meaning and that the entropy of a
black hole is one fourth of its horizon area.

\bigskip

These remarkable results of nearly three decades ago, point out to a deeper
relation among classical gravitation, quantum mechanics and statistical
properties. Hawking's semi-classical calculations allowed to interpret the
relation between classical black hole mechanics and its thermodynamics
leading to the celebrated entropy area expression.

\bigskip

During these years, different approaches have emerged to try to understand
the interplay between the quantum and classical descriptions, and the
statistical properties of black holes. We briefly mention the main quantum
gravity non-perturbative formalisms. String theory, whose building blocks
are essentially D-branes \cite{ref4}, loop quantum gravity \cite{ref5}, and
canonical quantum gravity, having as elementary constituents the quantum
excitations of the geometry itself . These proposals emerge essentially from
different principles. String theory is sought to be fundamental, it provides
the precise expression for the temperature and entropy. Loop and canonical
gravity do not provide the corresponding proportionality constants
unambiguously. These, however, deal directly with the curved black hole
geometry. In string theory, one carries out, for example, the calculation of
Hawking%
\'{}%
s radiation in flat space. Moreover, in string theory these calculations
have been up to now, only possible for extremal and nearly extremal black
holes \cite{ref4}.

\bigskip

In this paper, we are interested in canonical quantum gravity treatments
where a Hamiltonian quantum theory of spherically symmetric vacuum spacetime
can be defined \cite{ref6} (for our purposes we could also consider a ball
or shell of dust collapsing to a black hole \cite{ref7}). On the other hand,
supposing a uniformly spaced area spectrum \cite{ref8}, the Schwarzschild
black hole has been treated as a microcanonical ensemble \cite{ref9}. Also
considering a mini-superspace approach and by means of statistical
techniques previously used in hadron physics, the usual area of black hole
thermodynamics is recovered \cite{ref10}. Moreover, assuming that the area
spectrum of the black hole is uniformly spaced, a grand canonical ensemble
has been considered with the ADM mass (the Hamiltonian) and the horizon area
as separately observables \cite{ref11}. It is argued that in this way the
partition function is not divergent. However, the result in that work, a
logarithmic correction to the Bekenstein-Hawking entropy, is the same
already obtained by Kastrup \cite{ref12} by means of an analytic
continuation approach.

\bigskip

We will take into account the results mentioned in the previous paragraph,
and that the quantum Hamiltonians defined to study spherically symmetric
vacuum spacetime should describe, for an observer at rest far away, the
quantum mechanics of a black hole \cite{ref6,ref14}. We could have taken
also models of a shell or ball collapsing on its way to a black hole
formation \cite{ref7,ref15}. In our procedure it is not absolutely necessary
that the resulting area spectrum is uniformly spaced. However we need and
eigenvalue equation for the Arnowitt-Deser-Misner (ADM) mass of the hole or
its horizon area, which we do not consider independent variables. We will
proceed, with one of the most simple examples, by taking the Hamiltonian
operator defined by M\"{a}kela and Repo \cite{ref17}, taking $Q=0$, for the
Reissner- Nordstr\"{o}m black hole Hamiltonian. Our approach is
straightforward, an eigenvalue equation has been defined for a particular
parametrization of the variables of interest, where for a Schwarzschild
black hole a certain factor ordering has been chosen. The quantum-mechanical
model is represented by an eigenvalue equation for a linear harmonic
oscillator. For such quantum equations, we already know how to obtain the
corresponding statistical mechanics \cite{ref18}. For $Q\not=0$ and other
variable parametrizations and factor orderings, the same method outlined
here can be followed. We show how to proceed, with a simple example.

\bigskip

With Hamiltonians involving only momenta and coordinates, we can use path
integrals to get the corresponding statistical properties. The equation of
interest, in this paper, can be obtained from that in M\"{a}kela and Repo 
\cite{ref17}, for $Q=0$ is

\begin{equation}
\frac{\hbar ^{2}G^{2}}{c^{6}}a^{-s-1}\frac{d}{da}\left( a^{s}\frac{d}{da}%
\Psi (a)\right) =(a-\frac{2GM}{c^{2}})\Psi (a),
\end{equation}
where $a$ and $P_{a}^{2}=$ -$\frac{\hbar ^{2}G^{2}}{c^{6}}a^{-s}\frac{d}{da}%
\left( a^{s}\frac{d}{da}\Psi (a)\right) $ are phase coordinates obtained
from the phase space coordinates $m$ and $P_{m}$, by means of an appropriate
canonical transformation and $m(t):=M(t,r)$ and $p_{m}(t):=\int_{-\infty
}^{\infty }drP_{M}(t,r)$. The variable $m$ can be defined as the mass $M$ of
the hole when Einstein's equations are satisfied \cite{ref17}, $s$ is a
factor ordering parameter. In particular, if we choose $s=2$ and identifying 
$R_{s}=\frac{2GM}{c^{2}}$, one gets

\begin{equation}
\frac{\hbar ^{2}G^{2}}{c^{6}}\frac{1}{a}\left( \frac{d^{2}}{da^{2}}+\frac{2}{%
a}\frac{d}{da}\right) \Psi (a)=(a-R_{s})\Psi (a).
\end{equation}
We now transform

\begin{equation}
\Psi (a)=\frac{1}{a}U(a)\,\,{\rm and}\,\,x=a-R_{s},
\end{equation}
where the variable $x$ describes the gravitational degrees of freedom of the
Schwarzschild black hole. We also introduce the appropriate constants and
consider the fact that the energy of ``excitation'' associated with the
variable $a$ is not positive; the physical reason for this is simply that
the total energy of the black hole is included and the ADM energy is equal
to zero. Then, the quantum equation (2) transforms into

\begin{equation}
\left( -\frac{1}{2}\ell _{p\ell }^{2}E_{p\ell }\frac{d^{2}}{dx^{2}}+\frac{%
E_{p\ell }}{2\ell _{p\ell }^{2}}x^{2}\right) U(x)=\frac{R_{s}}{4\ell _{p\ell
}}E_{s}U(x),  \label{eq3}
\end{equation}
where $E_{s}=Mc^{2}$ is the black hole ADM energy and $\ell _{p\ell }=\sqrt{%
\frac{G\hbar }{c^{3}}}$ and $E_{p\ell }=\sqrt{\frac{c^{5}\hbar }{G}}$ are
Planck's length and energy respectively. Being (\ref{eq3}) a quantum linear
oscillator, one immediately gets

\begin{equation}
\frac{R_{s}(n)}{4\ell _{p\ell }}E_{s}(n)=\left( n+\frac{1}{2}\right)
E_{p\ell },  \label{eq5}
\end{equation}
this result coincides with Beckenstein's proposal \cite{ref1,ref8}.

Our procedure can be summarized as follows. We start with the Wheeler-DeWitt
equation for the Schwarzschild black hole (\ref{eq3}), we then use Feynman%
\'{}%
s path integrals approach to statistical mechanics. This will allow us to
calculate the free energy and the partition function. So that the
statistical properties of the black hole, temperature and entropy will be
deduced.

The classical partition function for the harmonic oscillator is given by

\begin{equation}
Z_{class}=\frac{1}{\beta \hbar \omega }.
\end{equation}
Applying the path integral formalism, the changes in the potential function
due to quantum mechanical effects can be included by means of the use of 
\cite{ref18}. The ``corrected'' potential for the oscillator is

\begin{equation}
V=\frac{m\omega ^{2}}{2}\left( x^{2}+\frac{\beta \hbar ^{2}}{12m}\right),
\end{equation}
which gives the partition function

\begin{equation}
Z_{approx}=\frac{e^{-\frac{(\beta \hbar \omega )^{2}}{24}}}{\beta \hbar
\omega }.  \label{eq6}
\end{equation}
It should be pointed out that this result is for the Euclidean space.

We now utilize this method, that has been so successful in quantum
mechanics, in the quantum gravity model of interest. Obviously, we do not
have a quantum theory of gravity, but we have a quantum harmonic oscillator
eigenvalue equation for the Schwarzschild black hole (\ref{eq3}), that
provides the expected energy eigenvalues (\ref{eq5}) \cite{ref1,ref8}. We
should note that the Schr\"{o}dinger equation (\ref{eq3}) for black holes,
corresponds to a quantum oscillator in Lorentzian space. Consequently, we
need to modify the partition function defined in (\ref{eq6}) to go from the
Lorentzian to the Euclidean world. After a rotation $\beta \rightarrow
-i\beta $, which is equivalent to a Wick rotation $t\rightarrow i\tau $, the
partition function has the following form

\begin{equation}
Z_{approx}^{\ast }=i\frac{e^{\frac{(\beta \hbar \omega )^{2}}{24}}}{\beta
\hbar \omega }.  \label{eq7}
\end{equation}
As already stated by Kastrup \cite{ref12}, this analytic continuation should
be used to get the thermodynamical properties of interest, so we have for
the internal energy

\begin{equation}
\overline{E}=-\frac{\partial \ln (%
\mathop{\rm Im}%
Z_{approx}^{\ast })}{\partial \beta }=-Mc^{2},  \label{eq8}
\end{equation}
which should be the internal gravitational energy of the black hole.
According to equation (\ref{eq3}) the frequency of our harmonic oscillator
results in $\hbar \omega =\sqrt{\frac{3}{2\pi }}E_{p\ell }$, so from
equations (\ref{eq7}) and (\ref{eq8}), we get

\begin{equation}
\frac{E_{p\ell}^{2}}{8\pi }\beta ^{2}-Mc^{2}\beta -1=0,
\end{equation}
the positive solution for this equation for the case $E_{p\ell}\ll Mc^{2}$
leads to

\begin{equation}
\beta =\frac{8\pi Mc^{2}}{E_{p\ell }^{2}}\left[ 1+\frac{1}{8\pi }\left( 
\frac{E_{p\ell }}{Mc^{2}}\right) ^{2}\right] =\beta _{H}\left[ 1+\frac{1}{%
\beta _{H}Mc^{2}}\right] ,  \label{eq10}
\end{equation}
which is Hawking%
\'{}%
s temperature $\beta _{H}=\frac{1}{kT_{H}}$, plus a small correction, that
is the same as that obtained by Kastrup \cite{ref12}. This calculation is
straightforward, that is because the quantum equation corresponds to a
linear oscillator, but even if we would have a more complicated equation, we
would only need to calculate a more difficult path integral. This would be
the case, for example, for the Reissner-Nordstr\"{o}m black hole $Q\not=0$ 
\cite{ref17}.

The entropy for the model, can be obtained again, by means of the rotation $%
(\beta \rightarrow -i\beta $, \ $k\rightarrow ik)$, we get

\begin{equation}
\mathop{\rm Im}%
\left( \frac{S}{ik}\right) =\left[ \ln \left( 
\mathop{\rm Im}%
Z_{approx}^{\ast }\right) +\beta \overline{E}\right] ,
\end{equation}
at $\beta \hbar \omega =\beta \sqrt{\frac{3}{2\pi }}E_{pl}$. By putting the
expression in terms of the partition function

\begin{equation}
\mathop{\rm Im}%
\left( \frac{S}{ik}\right) =\left[ \ln \left( 
\mathop{\rm Im}%
Z_{approx}^{\ast }\right) -\beta \frac{\partial \ln (%
\mathop{\rm Im}%
Z_{approx}^{\ast })}{\partial \beta }\right] ,
\end{equation}
and substituting equation (\ref{eq6}) we get

\begin{equation}
\frac{S}{k}=\frac{A_{s}}{4l_{p\ell }^{2}}\left[ 1+\frac{1}{8\pi }\frac{%
E_{p\ell }^{2}}{\left( Mc^{2}\right) ^{2}}\right] ^{2}+\frac{1}{2}\ln \left( 
\frac{A_{s}}{4l_{p\ell }^{2}}\left[ 1+\frac{1}{8\pi }\frac{E_{p\ell }^{2}}{%
\left( Mc^{2}\right) ^{2}}\right] ^{2}\right) +\frac{1}{2}\ln (24)-1,
\label{eq12}
\end{equation}
where $A_{s}=4\pi R_{s}^{2}$ is the area of the horizon.

In terms of the Bekenstein-Hawking relation $\frac{S_{BH}}{k}=\frac{A_{s}}{%
4l_{p\ell }^{2}}$ and ignoring terms of higher order, we finally obtain the
logarithmic correction as previously obtained by Kastrup \cite{ref12} and
recently by Gour \cite{ref11} using different procedures.

\begin{equation}
\frac{S}{k}=\frac{S_{BH}}{k}+\frac{1}{2}\ln \left( \frac{S_{BH}}{k}\right) .
\end{equation}

We should mention that it has already been shown in \cite{ref12} that the
black hole thermal fluctuations are small (and the same results follow in
our case). Consequently the thermal interactions with the heat bath are
small and, a canonical statistical treatment seems justified. The above
previous results would then be understood as to provide the conditions on
the heat bath \cite{york} if it is to be in thermal equilibrium with the
back hole.

In this paper, we have shown how the Feynman method to obtain the
statistical mechanics of a system from its fundamental quantum mechanics can
be applied to some quantum black hole models. For this purpose, as an
example, we have chosen one of the simplest \cite{ref17} Wheeler-DeWitt
equations proposed in the literature to describe the quantum mechanics of a
black hole. For any Hamiltonian in terms only of momenta and coordinates,
the path integral method provides us with a powerful tool to describe
physical properties of the system of interest. So, the procedure can be
generalized to a wide variety of Wheeler-DeWitt equations that have been or
could be proposed for black holes. One can consider models describing
collapsing dust or a shell on the process to form a black hole \cite
{ref7,ref15}, or those concerned with spherically symmetric vacuum spacetime 
\cite{ref6,ref14}. The proposals can be for black hole metrics including
charge, angular momentum and even other possible physical properties. Even
in these more general cases, the minisuperspace approach leaves out,
necessarily, degrees of freedom that are not considered in the quantum model
and for some physical purposes it will probably been unable to provide some
particular desired physical results that may require to take into account,
for example, particle interactions and (or) higher angular momentum modes 
\cite{susskind}. We have chosen an equation in minisuperspace (4) that has
been parametrized in such a way that it provides Bekenstein area spectrum
(5) \cite{ref1,ref8}. These area eigenvalues have been used directly, by
other authors \cite{ref11,ref12} to get Hawking%
\'{}%
s entropy and the logarithmic correction. For the purpose of getting the
temperature (12) and the black hole entropy and its logarithmic correction
(16), we have shown here that the minisuperspace approach, by means of
Feynman%
\'{}%
s method, gives the expected physical properties already obtained by other
means \cite{ref9,ref10,ref11,ref12}.Some other Hamiltonians for more general
metrics, with and without matter and their supersymmetric generalizations,
are under study and will be reported elsewhere.

\vskip 2truecm 
\centerline{\bf Acknowledgments}

We thank Carlos Martinez and Pedro Ludwig for several remarks. This work was
supported in part by CONACyT grant 28454E, and CONACyT-NFS grant.

\end{document}